# Coordinated Conservation Voltage Reduction and Network Topology Reconfiguration in Active Distribution Networks

Rida Fatima, Hassan Zahid Butt, and Xingpeng Li

*University of Houston, 4222 Martin Luther King Blvd, Houston, 77204, Texas, USA*

## Abstract

Distribution networks rely on conservation voltage reduction (CVR) and network topology reconfiguration (NTR) to manage demand, distributed energy resources (DERs), and tighter operating margins. Yet these strategies are typically deployed independently, despite coupled effects on feeder voltages, losses, and loading. This paper addresses that gap by developing a day-ahead mixed-integer second order cone programming (MISOCP) framework that jointly optimizes feeder head voltage control and network reconfiguration. The formulation incorporates AC branch flow constraints, ZIP load models, DERs, battery energy storage systems, and multi substation radiality. Using modified IEEE 33 bus and 123 bus systems, four configurations isolate the individual and coordinated effects of CVR and NTR. On the 33-bus system under mixed ZIP loading, NTR reduces active energy loss by 2.50%, while coordinated operation reduces it by 2.91% relative to SDN and produces a 2.46% CVR induced active load reduction relative to SDNTR. Under constant impedance loading, coordinated operation reduces active energy loss by 17.65% relative to SDN, while CVR reduces active load by 14.61% relative to SDNTR. DER portfolio scaling shows a nonmonotonic coordinated load reduction response, strongest at 20% penetration and decreasing at higher levels. On the 123-bus system, NTR reduces losses by 19.3%, while CVR reduces voltage dependent load energy by 2.3 to 2.4%. Under loading stress, SDN becomes infeasible at 120% loading, whereas CEDNTR remains feasible. Overall, the results demonstrate complementary, operating condition dependent benefits shaped by load composition, DER portfolio scaling, feeder topology, voltage margins, and source constraints, highlighting the value of joint optimization.

## Nomenclature

<u>Sets:</u>

| | |
|---|---|
| $S$ | Set of Substations |
| $N$ | Set of Nodes |
| $E$ | Set of Branches |
| $\delta^{+}(n)$ | Set of branches directed out of node $n$ |
| $\delta^{-}(n)$ | Set of branches directed into node $n$ |
| $E^{sw}$ | Set of switchable branches |
| $T$ | Set of scheduling intervals |
| $G$ | Set of dispatchable generators |
| $PV$ | Set of photovoltaic units |
| $B$ | Set of battery energy storage systems |

<u>Indices:</u>

| | |
|---|---|
| $s$ | Substation $s$, an element of $S$ |
| $n$ | Node $n$, an element of set $N$ |
| $e = (i, j)$ | Branch $e$ connecting nodes $i$ and $j$, $e \in E$ |
| $t$ | Time $t$, an element of set $T$ |
| $g$ | Generator $g$, an element of set $G$ |
| $p$ | Solar $p$, an element of set $PV$ |
| $b$ | Battery $b$, an element of set $B$ |

<u>Parameters:</u>

| | |
|---|---|
| $P_{n,t}^{D}$ | Nominal active power demand at node $n$ and time $t$ |
| $Q_{n,t}^{D}$ | Nominal reactive power demand at node $n$ and time $t$ |
| $V_s^{base}$ | Base feeder head voltage |
| $V_s^{CVR,min}$ | Minimum allowable feeder head voltage under CVR |
| $P_{sub,max}$ | Maximum allowable active power supplied by a substation |
| $Q_{sub,max}$ | Maximum allowable reactive power supplied by a substation |
| $r_{ij}$ | Resistance of branch $(i, j)$ |
| $x_{ij}$ | Reactance of branch $(i, j)$ |
| $V_n^{rated}$ | Rated voltage magnitude at node $n$ |
| $V_n^{min}$ | Minimum allowable nodal voltage magnitude |
| $V_n^{max}$ | Maximum allowable nodal voltage magnitude |
| $I_{ij}^{max}$ | Maximum allowable current magnitude of branch |
| $M_{ij}^{V}$ | Big parameter for the switch dependent voltage drop constraint |
| $\overline{S_{ij}}$ | Active and reactive branch power flow limit used for switch status coupling |
| $J_e^0$ | Initial branch status, equal to 1 if closed and 0 if open |
| $N_{sw}^{max}$ | Maximum number of switching actions over the scheduling horizon |
| $K_Z^P$ | Active load constant-impedance |
| $K_I^P$ | Active load constant-current |
| $K_p^P$ | Active load constant-power |
| $K_Z^Q$ | Reactive load constant-impedance |
| $K_I^Q$ | Reactive load constant-current |
| $K_P^Q$ | Reactive load constant-power |
| $P_g^{max}$ | Maximum active power output of generator $g$ |
| $P_g^{min}$ | Minimum active power output of generator $g$ |
| $Q_g^{max}$ | Maximum reactive power output of generator $g$ |
| $Q_g^{min}$ | Minimum reactive power output of generator $g$ |
| $P_{p,t}^{av}$ | Available active power from PV unit $p$ at time $t$ |
| $soc_b^{min}$ | Minimum allowable state of charge |
| $soc_b^{max}$ | Maximum allowable state of charge |
| $C_b^{duration}$ | Charging duration of BESS $b$ |
| $D_b^{duration}$ | Discharging duration of BESS $b$ |
| $\eta_b^{chg}$ | Charging efficiency |
| $\eta_b^{dchg}$ | Discharging efficiency |
| $E_b^{cap}$ | Energy capacity of BESS $b$ |
| $E_b^{init}$ | Initial stored energy of BESS $b$ |
| $Q_b^{max}$ | Maximum reactive power capability of BESS $b$ |
| $S_b^{max}$ | Apparent power rating of the BESS inverter |

<u>Variables:</u>

| | |
|---|---|
| $v_{n,t}$ | Squared voltage magnitude at node $n$ and time $t$ |
| $i_{ij,t}$ | Squared current magnitude on branch $(i, j)$ at time $t$ |
| $J_{e,t}$ | Binary status of branch $e$ at time $t$, equal to 1 if closed and 0 if open. |
| $F^{+}_{e,t}, F^{-}_{e,t}$ | Nonnegative auxiliary connectivity flows in the forward and reverse directions of branch $e$ |
| $a_{(i,j),t}$ | Switching action indicator for branch $e$ at time $t$ |
| $P^{Sub}_{s,t}$ | Active power supplied by substation $s$ at time $t$ |
| $Q^{Sub}_{s,t}$ | Reactive power supplied by substation $s$ at time $t$ |
| $P_{ij,t}$ | Power flow on branch $e$ at time $t$ |
| $Q_{ij,t}$ | Reactive power flow on branch $e$ at time $t$ |
| $P^{dchg}_{b,t}$ | Active Power discharged by battery $b$ at time $t$ |
| $Q^{dchg}_{b,t}$ | Reactive power discharged by battery $b$ at time $t$ |
| $P^{curt}_{p,t}$ | Curtailed solar power $p$ at time $t$ |
| $P^{chg}_{b,t}$ | Power used to charge the battery $b$ at charging time $t$ |
| $Q^{chg}_{b,t}$ | Reactive power absorbed by battery $b$ at time $t$ |
| $P^{gen}_{g,t}$ | Power produced by $g$ at time $t$ |
| $Q^{gen}_{g,t}$ | Reactive Power produced by $g$ at time $t$ |
| $c_{b,t}$ | Battery charging status $b$ at time $t$ (Binary) |
| $d_{b,t}$ | Battery discharging status $b$ at time $t$ (Binary) |
| $E_{b,t}$ | Stored energy of BESS $b$ at time $t$ |
| $\sigma_{s,t}$ | Nonnegative artificial connectivity supply from substation at time |
| Auxiliary expressions: | |
| $P^{l}_{n,t}$ | Voltage-dependent active ZIP demand at node $n$ and time $t$ |
| $Q^{l}_{n,t}$ | Voltage-dependent reactive ZIP demand at node $n$ and time $t$ |
| $P^{inj}_{n,t}$ | Net active power injection at node $n$ and time $t$ |
| $Q^{inj}_{n,t}$ | Net reactive power injection at node $n$ and time $t$ |
| $P^{PV}_{p,t}$ | Active power injected by PV unit $p$ at time $t$ |
| $\epsilon_{ij,t}$ | Conic relaxation residual for branch $(i,j)$ at time $t$ |
| $\epsilon^{max}$ | Maximum absolute conic relaxation residual |

## 1. Introduction

Distribution networks are experiencing increasing operational stress due to rising demand, longer feeder lengths, and growing penetration of distributed energy resources (DERs), particularly photovoltaic (PV) generation and battery energy storage systems (BESSs) [1]. High DER penetration can create overvoltage conditions, increase tap changer and regulator activity, and complicate power flow management, motivating the need for operational strategies that maintain voltage quality while improving network efficiency [2]-[5].

Among these strategies, network topology reconfiguration (NTR) and conservation voltage reduction (CVR) offer two distinct mechanisms for influencing feeder voltages, losses, and loading. NTR redistributes power flows by changing sectionalizing and tie-switch states, improving losses and loading without physical reinforcement [6]. CVR reduces voltage-dependent demand by operating feeder voltages near the lower end of the allowable range [7], [8].

Although both strategies have been widely studied, they are commonly optimized independently. This separation overlooks their direct physical interaction: CVR modifies nodal voltages and voltage-dependent demand, altering branch flows and the preferred feeder topology, while NTR reshapes electrical distances and voltage profiles, influencing how effectively CVR can reduce demand. Optimizing these controls separately can therefore miss operating points created by their interaction.

### *1.1 Network Topology Reconfiguration*

Network topology reconfiguration (NTR) improves distribution network performance by changing sectionalizing and tie-switch states to redistribute power flows, reduce losses, and alleviate loading constraints [6]. A major advance in this area came from Jabr [9], who showed that radial distribution load flow can be represented using conic programming, enabling computationally tractable optimization of distribution networks. Building on this foundation, Jabr et al. [10] and Taylor and Hover [11] developed mixed-integer second-order cone programming (MISOCP) models for minimum loss network reconfiguration under radiality constraints.

The conic representation adopted in this work follows the branch flow formulation of Farivar and Low [12], who established conditions under which the second-order cone relaxation is exact for radial networks. Gan et al. [13] further analyzed exact convex relaxation of optimal power flow in radial feeders, including standard test systems such as the IEEE 123-bus network. These studies provide the theoretical basis for using conic branch flow models in radial network optimization. However, because the present framework includes binary reconfiguration, multiple substations, voltage-dependent loads, DERs, and BESS operation, relaxation tightness is verified directly through post-solution residual analysis.

Subsequent research has expanded NTR to a wide range of operational objectives, including reliability improvement [14], time-varying operation and parallel computation [15], uncertainty-aware optimization [16], coordinated reactive power control [17], inverter-based voltage support [18], computationally efficient power flow formulations [19], dynamic reconfiguration [20], congestion management with renewable generation and storage [21], and increased distributed generation hosting capacity [22]. These studies demonstrate the broad operational value of topology reconfiguration. However, they primarily optimize network structure, power flow, reactive power, or DER operation in isolation, rather than coordinating topology decisions with feeder-voltage reduction and the resulting voltage-dependent demand response.

### *1.2 Conservation Voltage Reduction*

Conservation voltage reduction (CVR) lowers feeder demand by operating voltages near the lower end of the allowable range and exploiting the behavior of voltage-sensitive loads [7], [8]. Its effectiveness therefore depends strongly on load composition, particularly the constant-impedance and constant-current components that respond directly to voltage changes. Conventional CVR relies on on-load tap changers (OLTCs), capacitor banks, and voltage regulators [23], [24], while recent work has coordinated smart inverters with legacy voltage-control devices to support CVR in networks with increasing DER penetration [18], [25], [26].

Accurate CVR assessment requires voltage-dependent load modeling. The ZIP model represents demand using constant-impedance, constant-current, and constant-power components, allowing the present formulation to capture how load composition influences CVR performance and its interaction with NTR. Through this representation, voltage reductions at the feeder head translate into changes in nodal demand, branch flows, and losses, effects that become tightly coupled with the topology selected by NTR.

### *1.3 Coordinated CVR and NTR: Research Gap and Positioning*

Coordinated optimization of CVR and NTR remains comparatively limited. Closely related studies primarily combine network reconfiguration with Volt/VAR optimization, including MISOCP based reactive power coordination [17] and distributionally robust Volt/VAR control under PV uncertainty [27]. Other work coordinates CVR with inverter-based generation and soft open points but does not explicitly incorporate network reconfiguration [28]. This distinction is important because CVR and NTR influence the network through fundamentally different mechanisms: CVR alters voltage-dependent demand, branch currents, and losses, while NTR reshapes feeder paths, electrical distances, and downstream voltage constraints.

Optimizing these controls separately can therefore miss operating points created by their interaction.

This paper addresses that gap through a unified day-ahead MISOCP framework that jointly optimizes feeder-head voltage setpoints and binary reconfiguration while incorporating ZIP loads, DER scheduling, BESS operation, and multi-substation radiality. The coordinated value is evaluated across load compositions, DER portfolio scaling, and feeder sizes, revealing operating conditions that may not be captured when CVR and NTR are considered independently.

### *1.4 Contributions*

The main contributions of this work are:

1. A unified day ahead MISOCP framework that jointly optimizes CVR, NTR, DER dispatch, and BESS operation in multi substation radial networks with voltage dependent ZIP loads.
2. A controlled comparison of four operating configurations (SDN, CEDN, SDNTR, and CEDNTR) that isolates the individual and coordinated effects of CVR and NTR on losses, demand, voltage profiles, and feeder boundaries.
3. A systematic ZIP load sensitivity analysis demonstrating the strong dependence of CVR effectiveness on load voltage sensitivity, with coordinated CVR and NTR achieving up to 17.65% loss reduction and 14.61% CVR induced active load reduction under constant impedance loading.
4. A DER portfolio scaling sensitivity analysis showing that the coordinated load reduction benefit varies nonmonotonically with DER level, with the strongest active load reduction occurring at the investigated 20% reference scaling level.
5. A scalability and robustness demonstration on the IEEE 123 bus feeder, where NTR provides the dominant loss reduction effect and coordinated CVR-NTR remains feasible at 120% loading while the fixed topology baseline becomes infeasible.

The remainder of this paper is structured as follows. Section 2 presents the proposed MISOCP formulation. Sections 3 and 4 present case studies on the modified IEEE 33 bus and IEEE 123 bus systems, respectively. Section 5 summarizes the main findings and implications for distribution network operation.

## 2. Network Modeling & Optimization Formulation

The proposed framework minimizes active energy losses over a 24-hour scheduling horizon by jointly optimizing feeder-head voltage control, NTR, and DER operation in a radial distribution network. The formulation is based on an AC branch flow model with second-order cone relaxation and is expressed as a MISOCP. It incorporates voltage-dependent ZIP demand, DER and BESS scheduling, branch switching, multi-substation radiality, voltage and thermal limits, and switching constraints, enabling a unified representation of CVR, NTR, and resource operation within a single optimization framework.

### *2.1 Auxiliary Variable Definitions*

To obtain a conic-compatible branch flow formulation, squared voltage and squared current magnitude variables are introduced:

$$i_{ij,t} = |I_{ij,t}|^2 \, , \; v_{n,t} = |V_{n,t}|^2$$

Here, $v_{n,t}$ and $i_{ij,t}$ are optimization variables representing the squared nodal voltage magnitude and squared branch current magnitude, respectively. Using squared quantities avoids explicit square-root expressions in the branch flow equations and enables the voltage–current–power relationship to be represented through a second-order cone constraint. Voltage and current magnitudes required for reporting are recovered after optimization using:

$$\sqrt{i_{ij,t}} = I_{ij,t}\ ,\ \sqrt{v_{n,t}} = V_{n,t}$$

The squared-voltage variable $v_{n,t}$ is also used directly in the constant-impedance component of the ZIP load model, while the constant-current component is represented using a linear approximation of voltage magnitude, as described in Section 2.5.

### *2.2 Objective function*

The optimization objective minimizes total active power losses across all distribution lines over the 24-hour scheduling horizon:

$$min \sum_{ij \in E} \sum_{t \in T} r_{ij} i_{ij,t} \tag{1}$$

Where term $r_{ij} i_{ij,t}$ represents the active power loss on branch $(i,j)$ during interval $t$. Because each interval has a duration of one hour, the summed hourly losses are numerically equivalent to the total daily active energy loss, and the time step duration is omitted from (1).

The objective focuses exclusively on minimizing network losses rather than total supplied energy or customer demand. As a result, CVR is selected only when the associated voltage reduction contributes to the loss-minimizing operating point. Any reduction in voltage-dependent load is therefore treated as a performance outcome rather than a direct optimization target. Because the feeder head voltage is an optimization variable coupled to voltage dependent ZIP demand, its effect on branch currents and losses is inherently captured in the objective, allowing the optimizer to select the loss minimizing voltage without a separate CVR incentive term.

### *2.3 Power Flow and Network Constraints*

The active and reactive power balance at each node is enforced through (2) and (3). Equation (2) requires the net active power injection at each node to equal the difference between incoming and outgoing active power flows, adjusted for branch losses. Equation (3) imposes the corresponding reactive power balance. Together, these constraints ensure nodal power balance at every scheduling interval:

$$\textstyle\sum_{(k,i)\in E} (P_{ki,t} - r_{ki} i_{ki,t}) - \sum_{(i,j)\in E} P_{ij,t} + P_{i,t}^{inj} = 0, \qquad \forall i \in N,\ t \in T \tag{2}$$

$$\textstyle\sum_{(k,i)\in E} (Q_{ki,t} - x_{ki} i_{ki,t}) - \sum_{(i,j)\in E} Q_{ij,t} + Q_{i,t}^{inj} = 0, \qquad \forall i \in N,\ t \in T \tag{3}$$

Equation (4) represents the switch-dependent voltage-drop relation for each branch. When a branch is closed, the big-$M$ terms become inactive, and the constraint reduces to the standard branch flow voltage equation. When the branch is open, the big-$M$ terms relax the voltage coupling between its terminal nodes, allowing the topology to change without imposing a voltage relationship across disconnected branches:

$$-M_{ij}^{V}(1 - J_{ij,t}) \le v_{i,t} - v_{j,t} - 2(r_{ij} P_{ij,t} + x_{ij} Q_{ij,t}) + (r_{ij}^2 + x_{ij}^2) i_{ij,t} \le M_{ij}^{V}(1 - J_{ij,t}), \qquad \forall (i,j) \in E,\ t \in T \tag{4}$$

$$\left\| \begin{bmatrix} 2P_{ij,t} \\ 2Q_{ij,t} \\ i_{ij,t} - v_{i,t} \end{bmatrix} \right\|_2 \le i_{ij,t} + v_{i,t}, \qquad \forall (i,j) \in E\ ,\ t \in T \tag{5}$$

Equation (5) provides the second-order cone relaxation of the nonlinear voltage–current–power relationship. This constraint yields a tractable AC branch flow representation while preserving the coupling among active power, reactive power, squared current magnitude, and sending-end squared voltage. Relaxation tightness is evaluated after optimization using the validation metrics in Section 2.13.

### *2.4 Nodal Power Injection Modeling*

Net active and reactive power injections at each node are defined by (6) and (7). These expressions combine substation supply, PV generation after curtailment, BESS charging and discharging, dispatchable generation, and voltage-dependent load:

$$P_{n,t}^{inj} = \sum_{s\in S(n)} P_{s,t}^{Sub} + \sum_{p\in PV(n)}(P_{p,t}^{av} - P_{p,t}^{curt}) + \sum_{b\in B(n)}(P_{b,t}^{dchg} - P_{b,t}^{chg}) + \sum_{g\in G(n)} P_{g,t}^{gen} - P_{n,t}^{l} \tag{6}$$

$$Q_{n,t}^{inj} = \sum_{s\in S(n)} Q_{s,t}^{Sub} + \sum_{g\in G(n)} Q_{g,t}^{gen} + \sum_{b\in B(n)}(Q_{b,t}^{dchg} - Q_{b,t}^{chg}) - Q_{n,t}^{l} \tag{7}$$

These equations couple DER operation, BESS dispatch, PV curtailment, and voltage-dependent load to the nodal power balance constraints in (2)–(3), ensuring that all resource decisions directly influence network flows.

### *2.5 Voltage-Dependent Load Modeling*

Voltage-dependent demand is represented using the ZIP formulation, which models load as a combination of constant-impedance, constant-current, and constant-power components:

$$P_{n,t}^{l} = P_{n,t}^{D} * \left(K_Z^P * \frac{v_{n,t}}{(V_n^{rated})^2} + \frac{K_I^P}{2} * \left(\frac{v_{n,t}}{(V_n^{rated})^2} + 1\right) + K_p^P\right) \tag{8}$$

$$Q_{n,t}^{l} = Q_{n,t}^{D} * \left(K_Z^Q * \frac{v_{n,t}}{(V_n^{rated})^2} + \frac{K_I^Q}{2} * \left(\frac{v_{n,t}}{(V_n^{rated})^2} + 1\right) + K_P^Q\right) \tag{9}$$

$$K_Z^P + K_I^P + K_P^P = 1 \tag{10}$$

$$K_Z^Q + K_I^Q + K_P^Q = 1 \tag{11}$$

Equations (8) and (9) model active and reactive demand from nominal demand and the constant-impedance, constant-current, and constant-power coefficients, which satisfy (10) and (11). The impedance term varies with squared voltage, the power term is voltage-independent, and the current term is approximated using a first-order expression around nominal voltage. This approximation removes the square-root dependence while preserving the conic structure required for the MISOCP formulation.

### *2.6 Operational Constraints*

Network operating limits are enforced through (12)–(19):

$$\left(V_n^{min}\right)^2 \leq v_{n,t} \leq (V_n^{max})^2 \tag{12}$$

$$0 \leq i_{ij,t} \leq \left(I_{ij}^{max}\right)^2 J_{ij,t} \tag{13}$$

$$-S_{ij}J_{ij,t} \leq P_{ij,t} \leq S_{ij}J_{ij,t} \tag{14}$$

$$-S_{ij}J_{ij,t} \leq Q_{ij,t} \leq S_{ij}J_{ij,t} \tag{15}$$

$$P_g^{min} \leq P_{g,t}^{gen} \leq P_g^{max} \tag{16}$$

$$Q_g^{min} \leq Q_{g,t}^{gen} \leq Q_g^{max} \tag{17}$$

$$0 \leq P_{s,t}^{sub} \leq P_{sub,max} \tag{18}$$

$$0 \leq Q_{s,t}^{sub} \leq Q_{sub,max} \tag{19}$$

Constraint (12) maintains nodal voltages within allowable limits. Constraint (13) enforces thermal limits and forces branch current to zero when the branch is open. Constraints (14) and (15) couple branch power flows to switch status, preventing power transfer through open branches. Constraints (16) and (17) restrict generator outputs to their operating ranges. Finally, (18) and (19) limit substation injections according to available feeder-head capacity, preventing unrestricted upstream supply and defining the source capacity boundary for each feeder region.

## *2.7 Photovoltaic Curtailment Constraints*

PV curtailment is constrained by:

$$0 \leq P_{p,t}^{curt} \leq P_{p,t}^{av} \tag{20}$$

$$P_{p,t}^{PV} = P_{p,t}^{av} - P_{p,t}^{curt} \tag{21}$$

Constraint (20) limits curtailment between zero and the available PV generation at each interval, while (21) defines the actual active power injection as the available generation minus curtailed power. Together, these constraints ensure that PV output does not exceed the available resources and that curtailment remains physically valid.

## *2.8 Energy Storage System Modeling*

BESS operation is represented through state-of-charge limits, mutually exclusive charging and discharging modes, power limits, and intertemporal energy balance constraints. The scheduling horizon is defined as $T = \{1, \dots., N_T\}$

$$soc_{min}^{b} * E_{cap}^{b} \leq E_{b,t} \leq soc_{max}^{b} * E_{cap}^{b} \tag{22}$$

$$c_{b,t} + d_{b,t} \leq 1 \tag{23}$$

$$0 \leq P_{b,t}^{chg} \leq \frac{E_{cap}^{b}}{C_{b}^{duration}} c_{b,t} \tag{24}$$

$$0 \leq P_{b,t}^{dchg} \leq \frac{E_{cap}^{b}}{D_{b}^{duration}} d_{b,t} \tag{25}$$

$$E_{b,1} = E_{b}^{init} + \eta_{b}^{chg} P_{b,1}^{chg} - \frac{P_{b,1}^{dchg}}{\eta_{b}^{dchg}} \tag{26}$$

$$E_{b,t} = E_{b,t-1} + \eta_{b}^{chg} P_{b,t}^{chg} - \frac{P_{b,t}^{dchg}}{\eta_{b}^{dchg}}, \quad t = 2, \dots, N_T \tag{27}$$

$$E_{b,N_T} = E_{b}^{init} \tag{28}$$

$$0 \leq Q_{b,t}^{dchg} \leq d_{b,t}\, Q_{b}^{max} \tag{29}$$

$$0 \leq Q_{b,t}^{chg} \leq c_{b,t}\, Q_{b}^{max} \tag{30}$$

$$(P_{b,t}^{chg})^2 + (Q_{b,t}^{chg})^2 \leq c_{b,t}\, (S_{b}^{max})^2 \tag{31}$$

$$(P_{b,t}^{dchg})^2 + (Q_{b,t}^{dchg})^2 \leq d_{b,t}\, (S_{b}^{max})^2 \tag{32}$$

Constraint (22) enforces state-of-charge limits, while (23) prevents simultaneous charging and discharging. Constraints (24)– (25) impose power limits based on energy capacity and charge/discharge duration. Constraints (26)– (27) update stored energy using charging and discharging efficiencies, and (28) restores terminal energy to its initial value to prevent net depletion over the 24-hour horizon. Constraints (29)– (30) couple reactive power capability to the operating mode, and (31)–(32) enforce inverter apparent power limits during charging and discharging.

## *2.9 Network Topology Reconfiguration*

Network topology reconfiguration is represented using the binary branch status variable $J_{(i,j),t}$ where $J_{(i,j),t}$=1 indicates that branch $(i,j)$ is closed at time $t$, and $J_{(i,j),t}$=0 indicates that it is open. Only branches belonging to the switchable set $E^{sw}$ may change status; all remaining branches retain their initial configuration:

$$J_{ij,t} = J_{ij}^{0} \;\; (i,j) \notin E^{sw}, \forall t \tag{33}$$

Constraint (33) fixes all non-switchable branches at their initial state, while switchable branches may open or close when NTR is enabled. The branch current, active power, reactive power, and voltage drop constraints are all linked to $J_{ij,t}$. Consequently, an open branch carries no current or power and does not impose a voltage relationship between its terminal nodes. This representation allows the optimizer to modify feeder boundaries while preserving the physical consistency of the selected network topology.

### *2.10 Multi-Substation Radiality*

At each scheduling interval, the energized network must form a radial forest rooted at the substations. Radiality is enforced using auxiliary connectivity flow variables. The signed artificial connectivity flow on branch $(i, j)$ is defined as:

$$F_{(i,j),t} = F^{+}_{(i,j),t} - F^{-}_{(i,j),t} \tag{34}$$

where $F^{+}_{ij,t}$ and $F^{-}_{ij,t}$ are nonnegative forward and reverse flows. Connectivity flow is restricted to closed branches:

$$-(|N| - |S|)J_{(i,j),t} \le F_{(i,j),t} \le (|N| - |S|)J_{(i,j),t} \tag{35}$$

Radiality and connectivity are enforced through:

$$\begin{gathered} \sum_{e\in\delta^-(n)} F_{(i,j),t} - \sum_{e\in\delta^+(n)} F_{(i,j),t} = \begin{cases} 1, & n \in N \setminus S, \\ -\sigma_{s,t}, & n \in S, \end{cases} \\ \sum_{s\in\mathcal{S}} \sigma_{s,t} = |N| - |S|, \\ \sum_{e\in\mathcal{E}} J_{(i,j),t} = |N| - |S|, \quad \forall t \in T \end{gathered} \tag{36}$$

Equation (36) enforces nodal connectivity, total substation connectivity supply, and the branch count required to form a multi-substation radial forest. Together, these constraints ensure that every energized component is connected to exactly one substation while preserving radial operation.

Switching actions are limited by:

$$\begin{gathered} a_{ij,t} = |J_{ij,t} - J_{ij,t}{}^{-}|, \\ J_{ij,t}{}^{-} = \begin{cases} J^{0}_{(i,j)} & t = 1, \\ J_{ij,t-1}, & t > 1, \end{cases} \\ \sum_{t\in T} \sum_{e\in E^{\mathrm{sw}}} a_{(i,j),t} \le N^{\max}_{\mathrm{sw}} \end{gathered} \tag{37}$$

Constraint (37) counts changes in branch status and limits total switching actions. The absolute-value relation is implemented using standard mixed-integer linearization. The switching budget is four actions for NTR-enabled cases, and no terminal-topology constraint is imposed.

### *2.11 Conservation Voltage Reduction*

Conservation voltage reduction (CVR) is represented through control of the feeder-head voltage setpoint. When CVR is disabled, the substation voltage is fixed at its prescribed base value:

$$v_{s,t} = \left(V_s^{base}\right)^2 \tag{38}$$

When CVR is enabled, the feeder-head voltage is optimized continuously within its allowable range:

$$\left(V_s^{CVR,min}\right)^2 \le v_{s,t} \le \left(V_s^{base}\right)^2 \tag{39}$$

Through the ZIP load model, reducing the feeder-head voltage decreases the voltage-dependent portions of active and reactive demand. The substation voltage setpoint is therefore optimized jointly with DER dispatch, BESS operation, and network topology, while all nodal voltages remain subject to the limits defined in (12). The model assumes continuous feeder-head voltage control; discrete tap positions and detailed OLTC or regulator logic are not explicitly represented.

### *2.12 Coordinated Optimization Framework*

The proposed day-ahead MISOCP framework jointly optimizes network topology, feeder head voltage set points, substation power injections, DER dispatch, PV curtailment, and BESS operation to minimize total active energy losses. The optimization is subject to the branch flow model, voltage-dependent ZIP demand, network operating limits, BESS constraints, multi-substation radiality, and the prescribed switching budget.

CVR and NTR are optimized simultaneously. Network reconfiguration determines the feeder areas supplied by each substation, while feeder head voltage control influences voltage-dependent demand and the resulting branch flows and losses. Their coordinated optimization enables feeder boundaries, voltage setpoints, and resource schedules to adapt jointly over the 24-hour horizon.

### *2.13 Solution Strategy and Post-Solution Validation*

The proposed MISOCP model is implemented in Python using Pyomo and solved with Gurobi 11.0.3. The complete 24-hour scheduling horizon is optimized simultaneously with a one-hour resolution, allowing network topology, feeder-head voltage setpoints, DER dispatch, and BESS operation to be coordinated across all intervals. The IEEE 33 bus optimization cases were solved using a relative MIP gap tolerance of 0.1%.

Post-solution checks verify network feasibility and the tightness of the conic relaxation. The selected topology is reconstructed for each interval to confirm radiality, substation connectivity, absence of isolated load buses, and compliance with the switching limit. Power balance, voltage drop consistency, open branch flow, and voltage limit residuals are also evaluated.

The second order cone relaxation is validated using the residual

$$\in_{ij,t} = i_{ij,t} v_{i,t} - P_{ij,t}^2 - Q_{ij,t}^2 \qquad (40)$$

and the maximum residual across all closed branches and intervals:

$$\in^{max} = \max_{(i,j),t} | \in_{ij,t} | \qquad (41)$$

Small residual values indicate that the relaxed solution closely satisfies the original nonlinear voltage–current–power relationship.

### *2.14 Evaluation Scenarios and Performance Metrics*

Four operating configurations are evaluated to isolate the individual and coordinated effects of CVR and NTR. All cases use identical network data, load and DER profiles, operating limits, BESS assumptions, and scheduling conditions. The configurations differ only in whether CVR and NTR are enabled, as summarized in Table 1.

The configurations are compared using total active energy loss, active load energy, minimum nodal voltage, switching activity, and inter-substation load transfer. Network efficiency is evaluated using total active energy loss and its percentage reduction relative to SDN. Because CVR affects voltage-dependent demand, active load energy and the corresponding load reduction are also reported. Voltage performance is assessed using the minimum nodal voltage, while NTR performance is evaluated through switching actions and load transfer between substation service areas.

Table 1. Evaluated Operating Configurations

| *Case* | *CVR* | *NTR* | *Network operation* |
|---|---|---|---|
| SDN | Disabled | Disabled | Fixed topology and fixed substation voltages |
| SDNTR | Disabled | Enabled | Network reconfiguration with fixed substation voltages |
| CEDN | Enabled | Disabled | Fixed topology with optimized substation voltage control |
| CEDNTR | Enabled | Enabled | Coordinated voltage control and network reconfiguration |

## 3. Case Study and Performance Evaluation

The proposed framework is first evaluated on a modified IEEE 33 bus distribution system to examine the individual and coordinated effects of CVR and NTR. The IEEE 123 bus system is subsequently used to assess scalability and loading robustness. All configurations are evaluated over a 24-hour scheduling horizon under identical operating assumptions within each test system.

### *3.1 Modified IEEE 33-Bus Network Configuration*

The modified IEEE 33 bus system shown in Fig. 1 is based on the standard IEEE 33 bus feeder [6] and includes two independently controlled substations located at buses 1 and 33. The network contains 33 buses and 35 candidate branches. At each scheduling interval, 31 branches remain closed, forming two radial feeder regions, each connected to one substation. This configuration enables feeder boundary reconfiguration, inter substation load transfer, and coordinated voltage control.

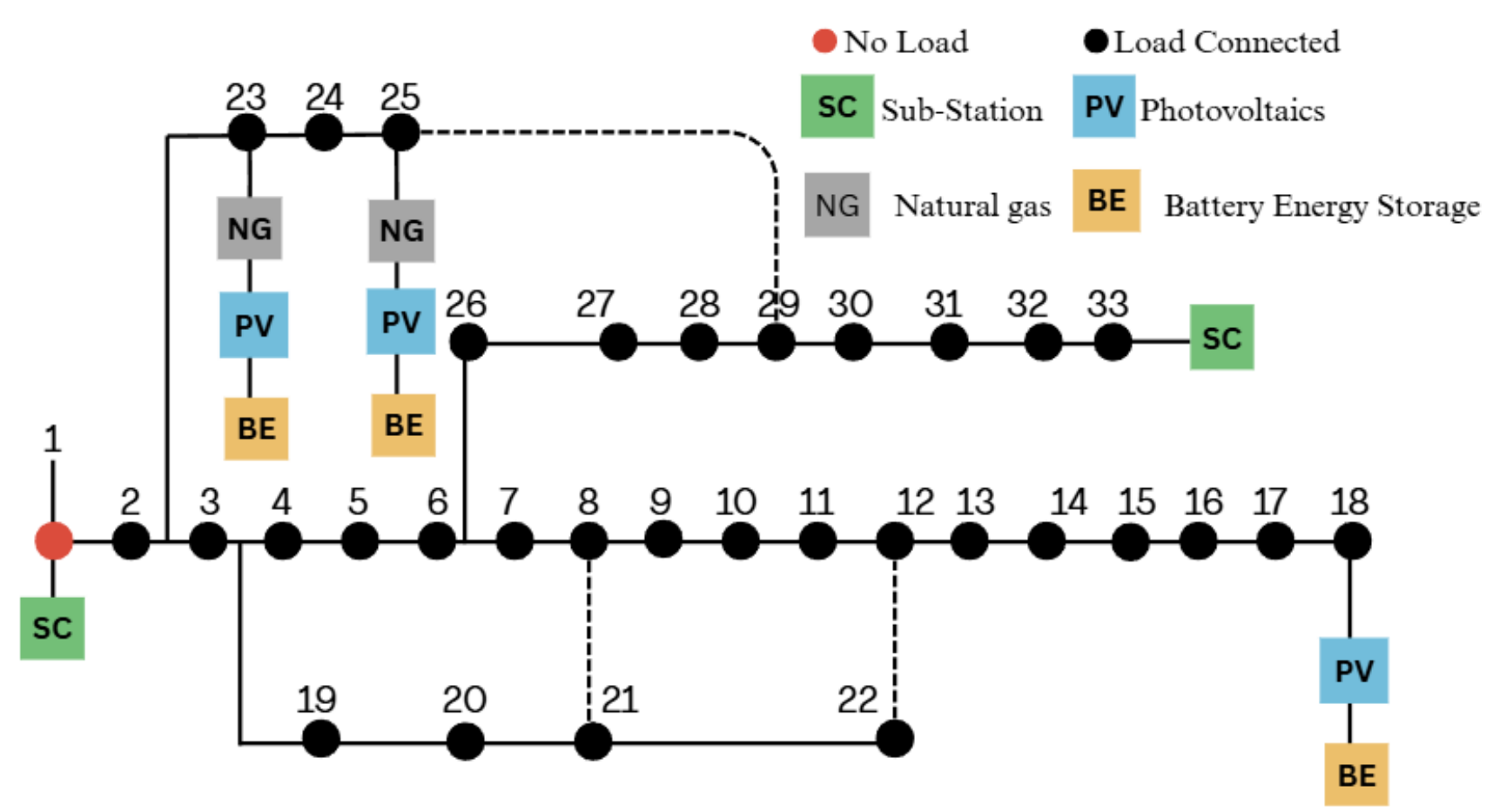

Fig. 1. Modified IEEE 33-bus system.

The base topology is used for the SDN and CEDN configurations, while designated switchable branches may change status in the SDNTR and CEDNTR configurations subject to the radiality and switching constraints described in Section 2. The locations of DERs and BESS units are also shown in Fig. 1. The network includes photovoltaic generation, BESSs, and dispatchable generators. Each substation is modeled as a controllable feeder head source subject to prescribed active and reactive power injection limits. When CVR is disabled, the substation voltage is fixed at 1.05 p.u. When CVR is enabled, each substation voltage setpoint is independently optimized within its prescribed range while all nodal voltages are maintained between 0.95 and 1.05 p.u. Upstream transmission network constraints beyond the specified substation injection limits are not explicitly represented.

Hourly variations in nominal demand are represented using the normalized 24-hour profile shown in Fig. 2. The complete horizon is optimized simultaneously using a one-hour resolution.

For the representative base case, both active and reactive demands use the mixed ZIP composition.

$$K_z = 0.24, K_i = 0.36, K_p = 0.40.$$

The same ZIP composition is applied at all load buses, while the hourly profile scales the nominal nodal demands. Alternative ZIP compositions are subsequently examined in the sensitivity analysis.

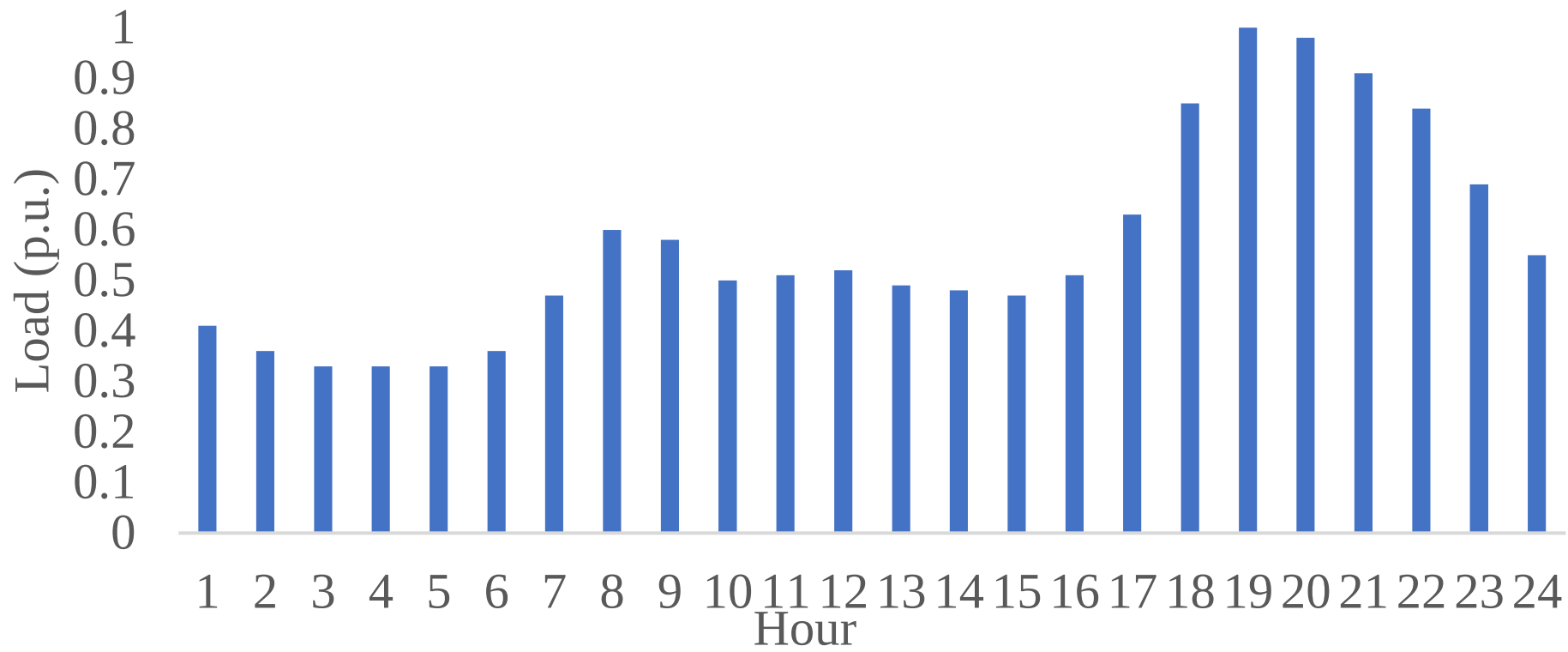

Fig. 2. Normalized 24-hour base load profile.

### *3.2 Base Case Comparative Assessment*

The base case compares all four configurations under the representative mixed ZIP composition using identical network, load, DER, voltage, and resource assumptions. Relative to SDN, active energy loss decreases by approximately 0.12% for CEDN, 2.50% for SDNTR, and 2.91% for CEDNTR, as shown in Fig. 3. The small improvement under CEDN indicates limited CVR benefit under the fixed topology, while NTR provides the dominant loss reduction. Coordinated CEDNTR achieves the lowest loss, showing an additional benefit when voltage control and reconfiguration are optimized jointly.

Table 2 further shows that CEDNTR reduces average active load from approximately 2.193 MW in SDN to 2.140 MW and average active loss from 0.0172 MW to 0.0167 MW. The peak substation import also decreases from approximately 3.522 MW to 3.415 MW. CEDNTR operates at a mean bus voltage of approximately 1.0163 p.u. and reaches a minimum voltage of 0.9500 p.u., while remaining within the prescribed voltage limits. Both NTR cases require only two switching operations. These results show that reconfiguration changes the loss minimizing operating point and enables a stronger CVR response under coordinated operation.

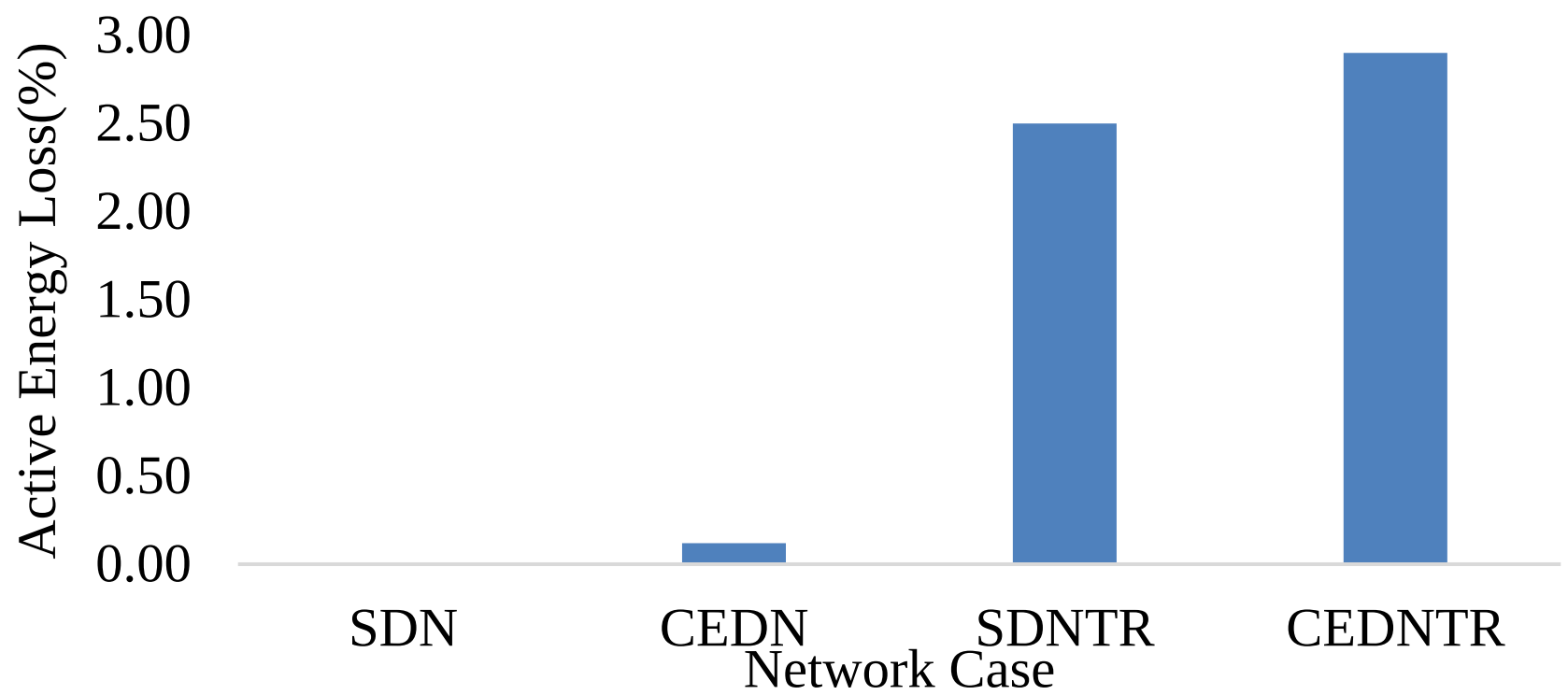

Fig. 3. Active energy loss reduction relative to SDN for the four operating cases.

Table 2. Base-Case performance of the evaluated configurations

| Case | Average Load (MW) | Average Loss **(**MW) | Mean Bus Voltage (p.u) | $\mathbf{V_{min}}$ (p.u) | Peak Import Substation (MW) |
|---|---|---|---|---|---|
| SDN | 2.193 | 0.01721 | 1.0403 | 1.0076 | 3.522 |
| CEDN | 2.193 | 0.01719 | 1.0401 | 1.0075 | 3.522 |
| SDNTR | 2.194 | 0.01678 | 1.0402 | 1.0077 | 3.508 |
| CEDNTR | 2.140 | 0.01671 | 1.0163 | 0.950 | 3.415 |

### *3.3 Mechanism and Interaction of CVR and Network Reconfiguration*

Fig. 4 compares the hourly network mean voltage under CEDN and CEDNTR. Although both configurations have the same feeder head voltage control range, CEDN maintains a relatively high mean voltage of approximately 1.0401 p.u. Under the fixed topology, substantial voltage reduction is therefore not selected by the loss minimizing solution. With network reconfiguration, CEDNTR operates at a considerably lower mean voltage of approximately 1.0163 p.u. Reconfiguration changes feeder paths, current distribution, and the resulting loss and voltage relationship, making lower voltage operation more favorable under coordinated operation.

Fig. 5 compares the mean and minimum nodal voltages under SDNTR and CEDNTR. Enabling CVR reduces the mean voltage from approximately 1.0402 p.u. in SDNTR to 1.0163 p.u. in CEDNTR, while the minimum voltage decreases from approximately 1.0077 p.u. to 0.9500 p.u. without violating the prescribed voltage limit. This change produces a 2.46% CVR induced reduction in active load relative to SDNTR. In contrast, CVR alone reduces active load by only about 0.01% under the fixed topology.

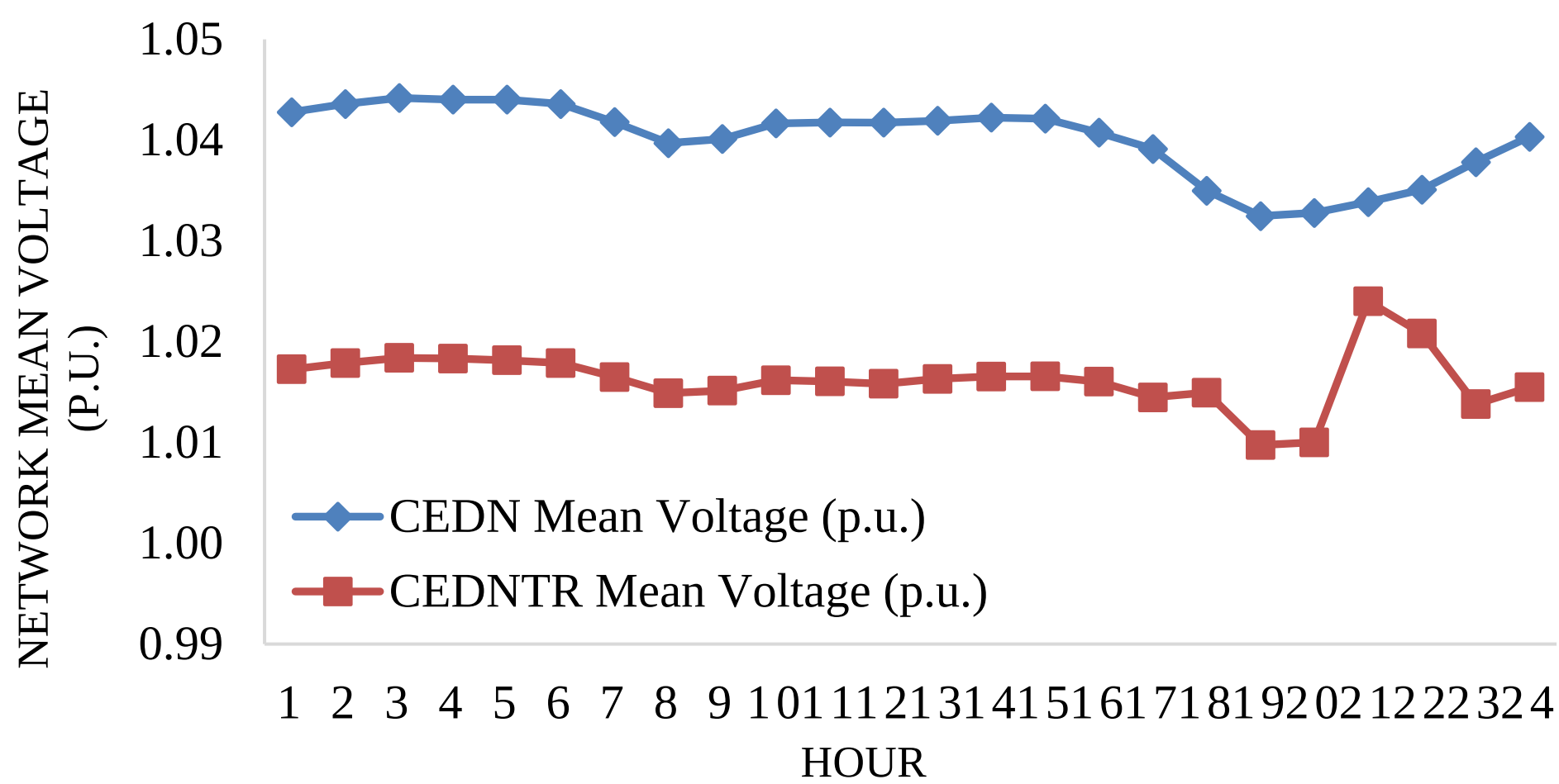

Fig. 4. Hourly network-mean voltage for CEDN and CEDNTR.

These results demonstrate the interaction between the two controls. NTR changes feeder paths and voltage conditions, while CVR modifies voltage dependent demand and the resulting power flows. In the representative mixed ZIP case, reconfiguration creates an operating point that allows CVR to use substantially more of the available voltage margin, resulting in a stronger demand reduction under coordinated operation.

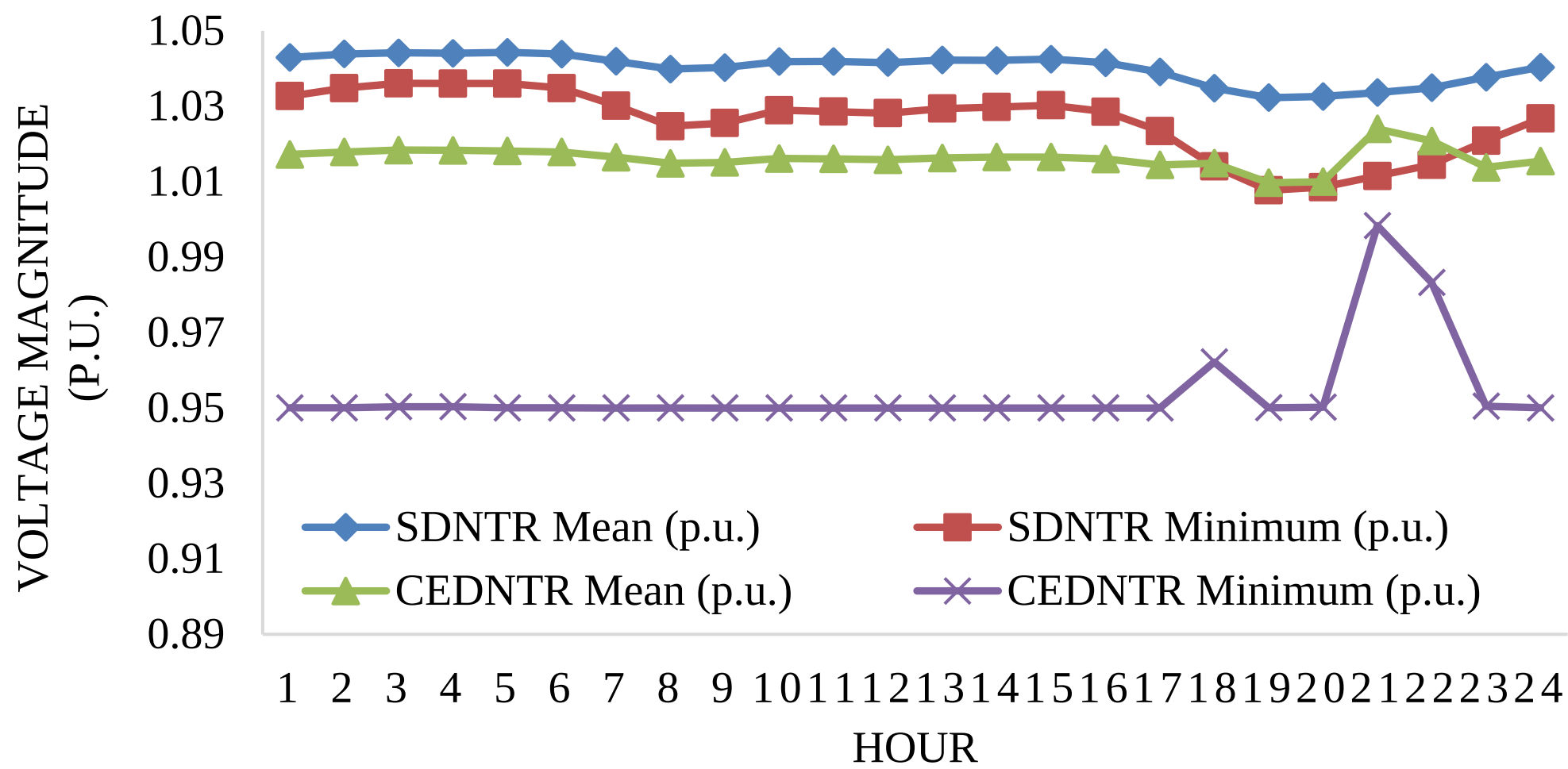

Fig. 5. Hourly mean and minimum bus voltages for SDNTR and CEDNTR.

### *3.4 Sensitivity to ZIP Load Composition*

The influence of load voltage sensitivity is evaluated by progressively increasing the constant-impedance share while reducing the constant-power and constant-current components. The representative mixed-ZIP load model uses $K_Z = 0.24$, $K_I = 0.36$, and $K_P = 0.4$. To systematically evaluate the effect of increasing voltage sensitivity, the constant-impedance coefficient $K_Z$ is progressively increased, while the remaining load fraction is allocated between the constant-current and constant-power components according to their baseline ratio. Since $K_I : K_P = 0.36 : 0.40 = 9:10$, the intermediate coefficients are determined as $K_I = \left(\frac{9}{19}\right)(1 - K_Z)$ and $K_P = (10/19)(1 - K_Z)$. This produces the evaluated ZIP compositions (0.24,0.36,0.40), (0.40,0.28,0.32), (0.60,0.19,0.21) and (0.80,0.09,0.11), together with the constant-current (0,1,0) and constant-impedance (1,0,0) boundary cases. A constant-power (0,0,1) boundary case is also evaluated separately.

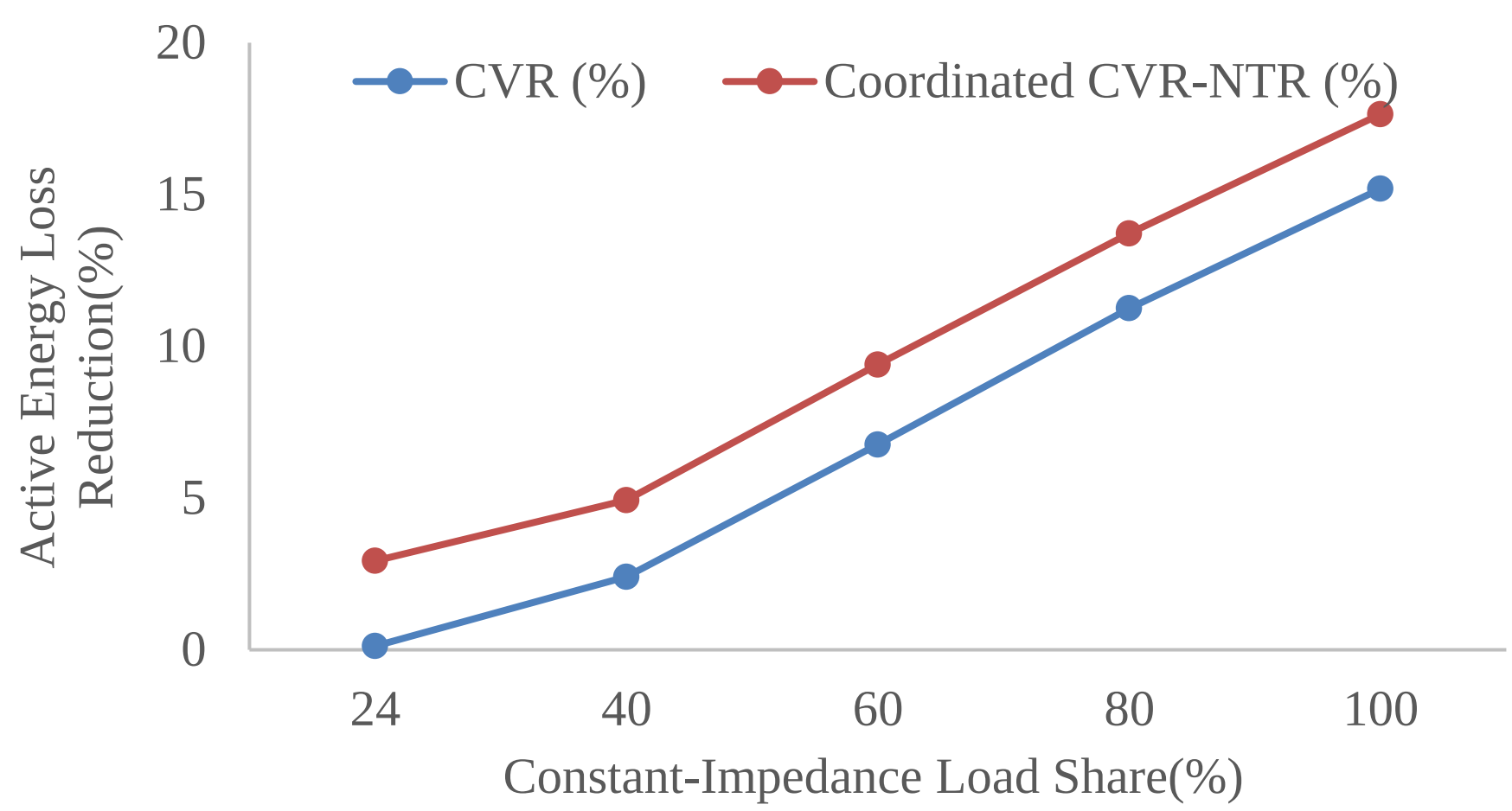

Fig. 6. Active energy loss reduction versus constant-impedance load share.

Fig. 6 shows that CVR effectiveness increases substantially as load voltage sensitivity increases. Under the representative mixed ZIP case, the CVR only loss reduction is approximately 0.12%, while coordinated CEDNTR reduces losses by approximately 2.91% relative to SDN**.** As the constant impedance share increases, the coordinated loss reduction rises to approximately 4.93% at 40%, 9.39% at 60%, 13.72% at 80%, and 17.65% relative to SDN under constant impedance loading. The corresponding CVR only reductions also increase progressively, confirming the strong dependence of CVR effectiveness on load composition.

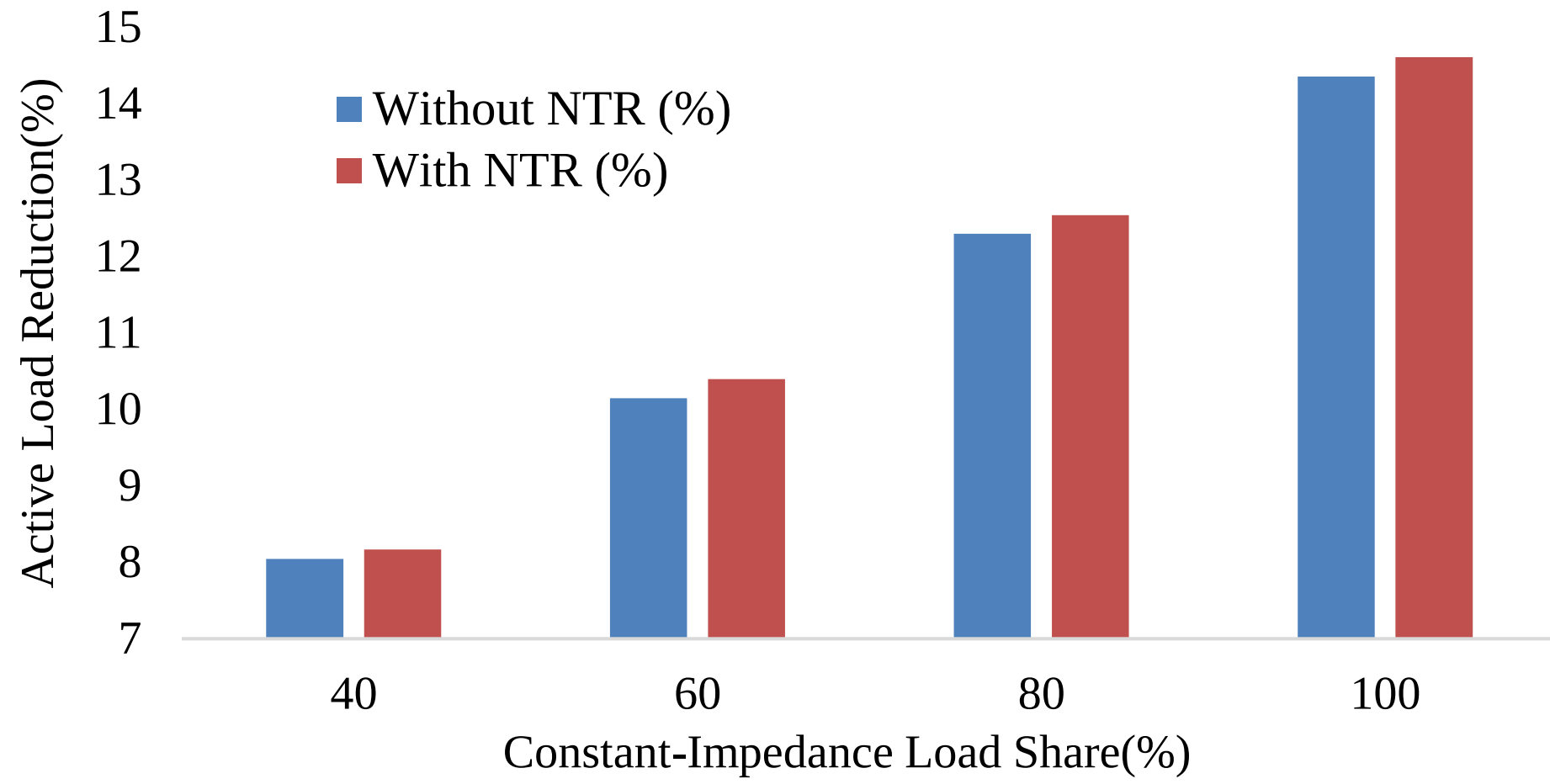

Fig. 7. Active load reduction versus constant-impedance load share.

For the additional ZIP sensitivity cases shown in Fig. 7, the constant impedance share is varied from 40% to 100%. Without NTR, CVR reduces active load by approximately 0.01% under the representative mixed ZIP case, increasing to 8.05%, 10.15%, 12.30%, and 14.36% as the constant impedance share rises from 40% to 100%. With NTR, the corresponding reductions are approximately 2.46%, 8.17%, 10.40%, 12.54%, and 14.61%. The largest coordination effect occurs under the representative mixed ZIP condition, where reconfiguration creates a substantially more favorable voltage operating point for CVR. As voltage sensitivity increases, CVR becomes increasingly effective even under the fixed topology, reducing the relative contribution of reconfiguration to demand reduction.

Constant power and constant current loads are also evaluated as boundary cases, as summarized in Table 3. Under constant power loading, CVR produces zero active load reduction, as expected because demand is independent of voltage magnitude. The small loss difference between SDN and CEDN is within numerical tolerance and is therefore not interpreted as a meaningful change. Under constant current loading, CVR reduces active load by approximately 7.19% without NTR and 6.28% with NTR, while CEDN and CEDNTR reduce losses by approximately 1.22% and 3.52%, respectively. These boundary cases confirm that the modeled CVR response is consistent with the expected ZIP load behavior.

Table 3. ZIP Boundary Case Results

| Load Case | CEDN loss reduction (%) | CEDNTR loss reduction (%) | Load reduction without NTR (%) | Load reduction with NTR (%) |
|---|---|---|---|---|
| Constant Power | 0.00 | 2.37 | 0 | 0 |
| Constant Current | 1.22 | 3.52 | 7.19 | 6.28 |

### *3.5 Sensitivity to DER Portfolio Scaling*

The effect of DER portfolio scaling is evaluated under the representative mixed ZIP load composition. The 20% case is used as the reference portfolio, with the 0%, 40%, and 80% cases representing zero, two times, and four times the reference DER level, respectively. PV generation, dispatchable generator ratings, and BESS power and energy capacities are scaled proportionally, while DER locations, technology mix, storage duration, load composition, and network parameters remain unchanged. As shown in Fig. 8, the normalized active energy loss ratio generally decreases with increasing DER scaling. NTR based configurations consistently outperform the corresponding fixed topology cases, while the CVR only contribution remains comparatively small. CEDNTR yields the lowest normalized loss ratio at all evaluated levels, with loss reductions relative to SDN of approximately 2.51%, 2.91%, 2.01%, and 2.70% for the 0%, 20%, 40%, and 80% cases, respectively.

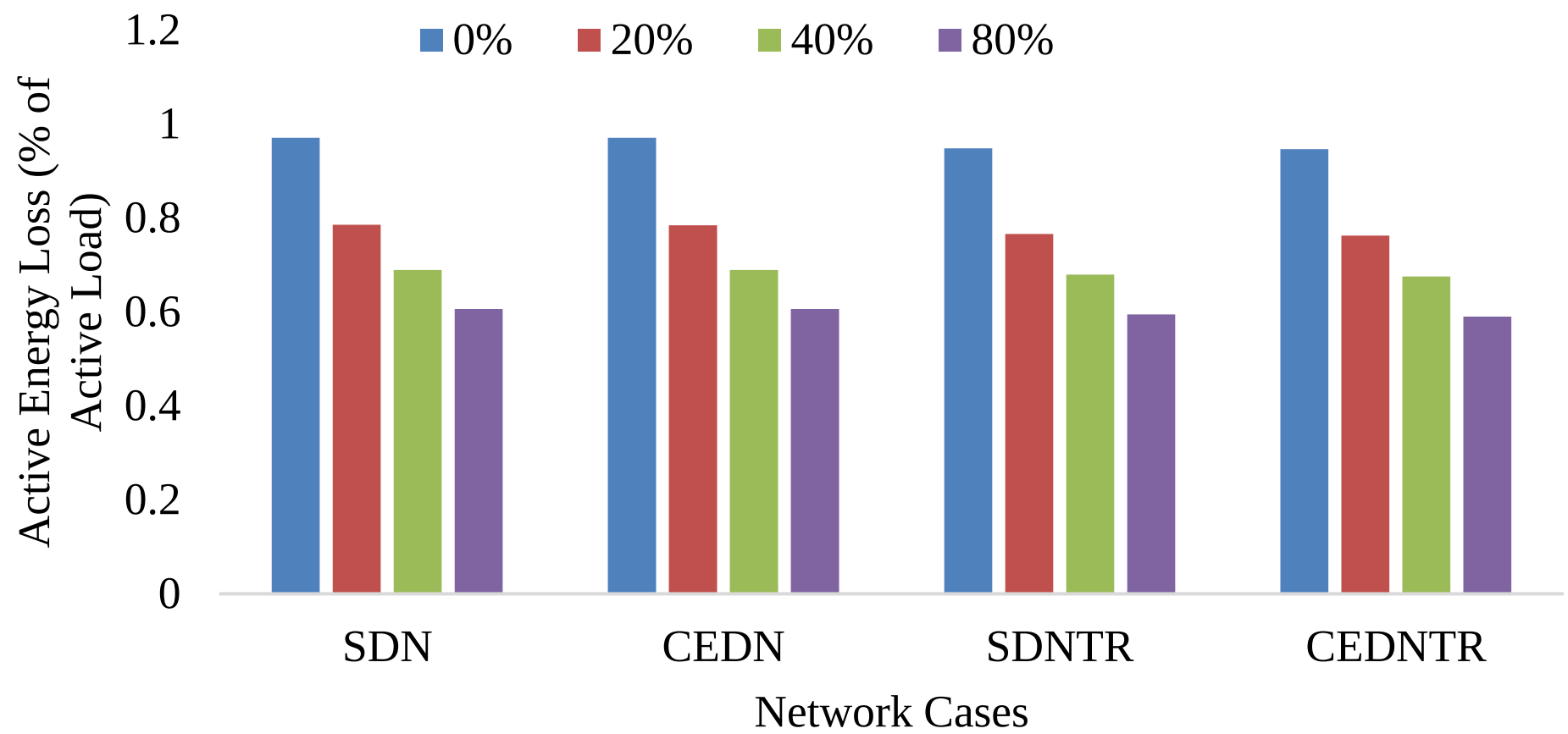

Fig. 8. Normalized active energy loss ratio across the evaluated DER portfolio scaling levels.

Fig. 9 shows the active load reduction achieved by CEDNTR relative to SDN. The reduction is negligible at 0% DER portfolio scaling, increases to approximately 2.45% at 20%, decreases to 2.07% at 40%, and falls to approximately 0.16% at 80%. This nonmonotonic response follows the optimized voltage operating point. CEDNTR operates close to the 0.95 p.u. lower voltage limit at 20% and 40% penetration, whereas the mean voltage increases substantially at 80%, reducing the achievable voltage dependent demand reduction.

These results demonstrate that increasing DER portfolio scaling does not produce a proportional increase in coordinated load reduction benefit. Instead, the benefit depends on the interaction among DER injections, feeder voltage conditions, network topology, and voltage dependent demand. The strongest combined CEDNTR load reduction response occurs at the investigated 20% DER portfolio scaling.

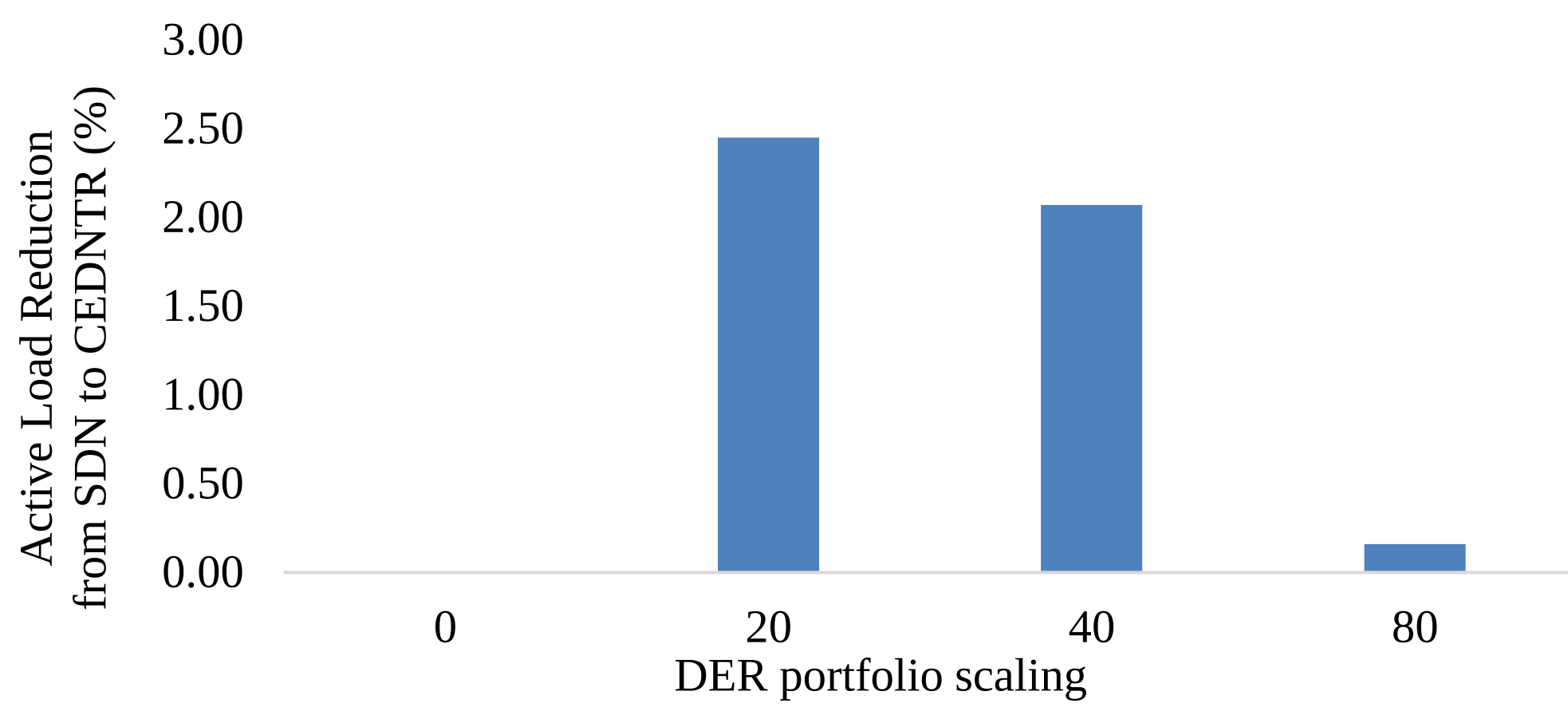

Fig. 9. Load reduction of CEDNTR relative to SDN baseline at various DER portfolio scaling.

## 4. Scalability and Loading Robustness on the IEEE 123 Bus System

To evaluate scalability and cross-feeder applicability, the proposed framework is further tested on a modified IEEE 123-bus distribution feeder [29]. The original unbalanced feeder is represented using a balanced equivalent to maintain consistency with the branch flow formulation used in the IEEE 33-bus study. The same 24-hour scheduling horizon, representative mixed ZIP load model, voltage limits, and DER assumptions are retained.

The modified system contains three substations located at buses 48, 65, and 76, and is operated as three substation-rooted radial regions. The resulting optimization problem contains approximately 27,288 variables, including 3,168 binary variables, 58,759 linear constraints, and 2,880 conic constraints. This larger network therefore provides a more demanding test of both the computational scalability and the operational robustness of the proposed formulation.

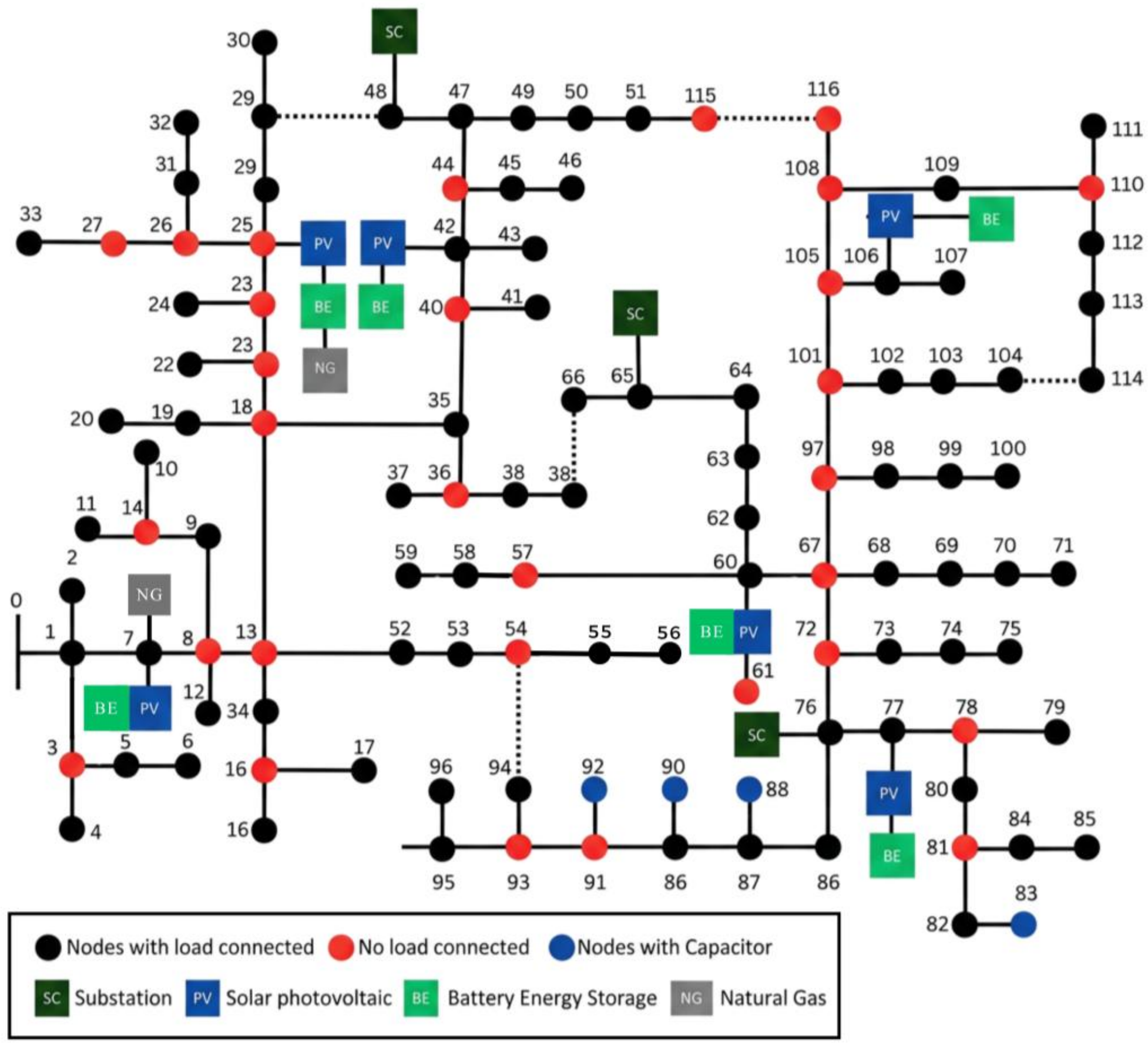

Fig. 10. IEEE Modified 123 bus system.

### 4.1 *Active Energy Loss and Voltage-Dependent Demand*

Table 4 compares the four IEEE 123-bus operating configurations. NTR produces a substantial reduction in daily active energy loss relative to the fixed topology SDN case. The SDNTR and CEDNTR cases yield very similar loss levels. The difference between their incumbent objective values is of the same order as the achieved MIP optimality gaps; therefore, the two solutions are not ranked based on this difference. The relatively small incremental loss benefit of CVR is consistent with the adopted ZIP load composition, which contains a modest constant impedance component ($K_z$=0.24), together with the operating conditions of the reconfigured feeder. Consequently, NTR provides the dominant loss reduction contribution on the larger feeder by redistributing branch power flows, whereas the principal contribution of CVR under this load mix is the reduction of voltage dependent demand rather than a substantial additional decrease in feeder losses. Fig. 11 shows that the fixed voltage SDN and SDNTR cases maintain comparatively higher minimum nodal voltages, whereas the CVR enabled CEDN and CEDNTR cases operate closer to the lower allowable voltage limit and approach 0.95 p.u. during high demand periods. Accordingly, changes in feeder losses should not be interpreted as equivalent changes in feeder load energy. Under the ZIP load model, active demand varies with the operating voltage; therefore, CVR can reduce load energy even when its incremental effect on network losses is small.

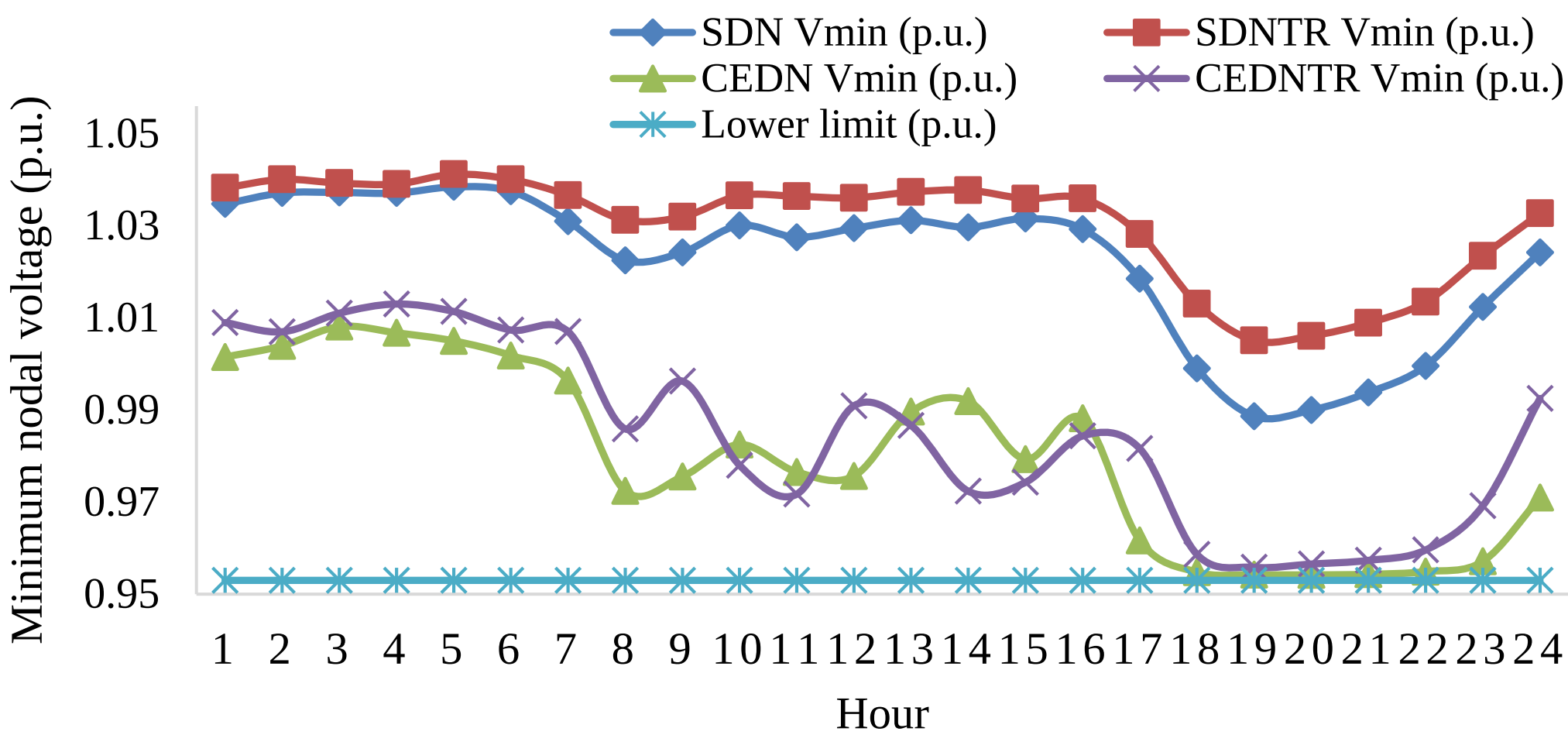

Fig. 11. Hourly minimum nodal voltage for the four IEEE 123-bus operating configurations.

Table 4. Performance Comparison for the Modified IEEE 123-Bus System

| Configuration | Loss (MWh) | Load Energy (MWh) | Vmin(p.u.) |
|---|---|---|---|
| SDN | 0.323 | 49.3 | 0.98 |
| SDNTR | 0.261 | 49.5 | 1.00 |
| CEDN | 0.321 | 48.1 | 0.95 |
| CEDNTR | 0.259 | 48.3 | 0.95 |

### *4.2 Computational and Numerical Performance*

The four IEEE 123-bus cases are solved in approximately 206–1,013 s using Gurobi 11.0.3, demonstrating that the proposed MISOCP formulation remains computationally tractable for the larger feeder. The maximum active and reactive power balance residuals are on the order of $10^{-9}$ and $10^{-11}$, respectively. The maximum positive conic-relaxation residual is on the order of $10^{-7}$, and the largest relative residual occurs on a branch carrying negligible power.

Radiality, substation connectivity, voltage limits, switching constraints, and open branch flow conditions are satisfied in all solved configurations. These results confirm the numerical consistency of the solutions obtained and support the applicability of the proposed formulation to the larger IEEE 123-bus system.

### *4.3 Loading Stress Assessment*

To evaluate loading robustness, the IEEE 123-bus demand is uniformly increased while maintaining the original network and operating limits. The fixed-topology SDN case remains feasible through 110% loading but becomes infeasible at 120%, whereas CEDNTR remains feasible at the tested 120% loading level. Thus, relative to the SDN baseline, CEDNTR demonstrates feasibility at a higher tested loading level under the investigated stress conditions.

Fig. 12 compares the peak active power utilization of the three substations. At 110% loading, the SDN case utilizes approximately 67.2%, 90.6%, and 92.6% of the capacities of Substations 1, 2, and 3, respectively. Branch thermal loading remains below its limit, with the most heavily loaded branch reaching approximately 72.5%. A diagnostic relaxation of only the Substation 3 active power limit restores SDN feasibility at 120% loading, confirming that the observed loading boundary is governed by feeder-head capacity rather than branch thermal limits.

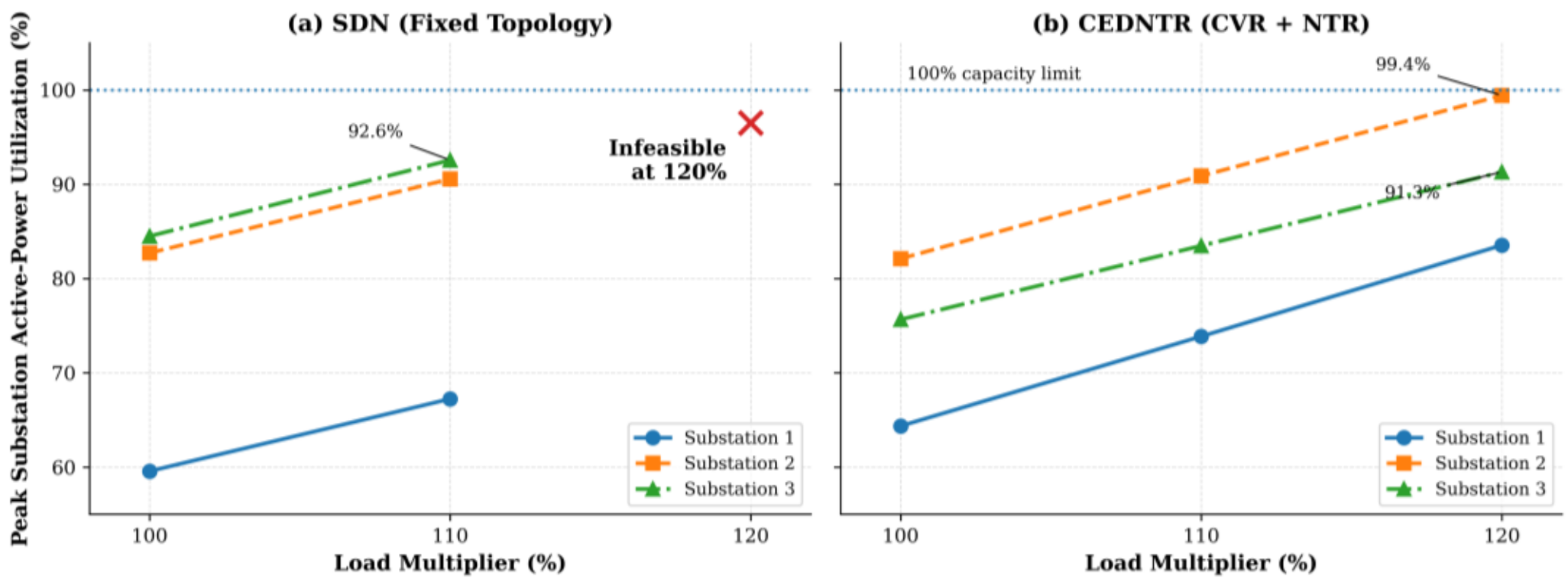

Fig. 12. Peak substation utilization under increasing load for SDN and CEDNTR.

Under CEDNTR at 120% loading, the peak utilizations of Substations 1, 2, and 3 are approximately 83.5%, 99.4%, and 91.3%, respectively. Network reconfiguration and voltage control redistribute feeder demand among the available substations, allowing the system to remain feasible under the higher loading condition.

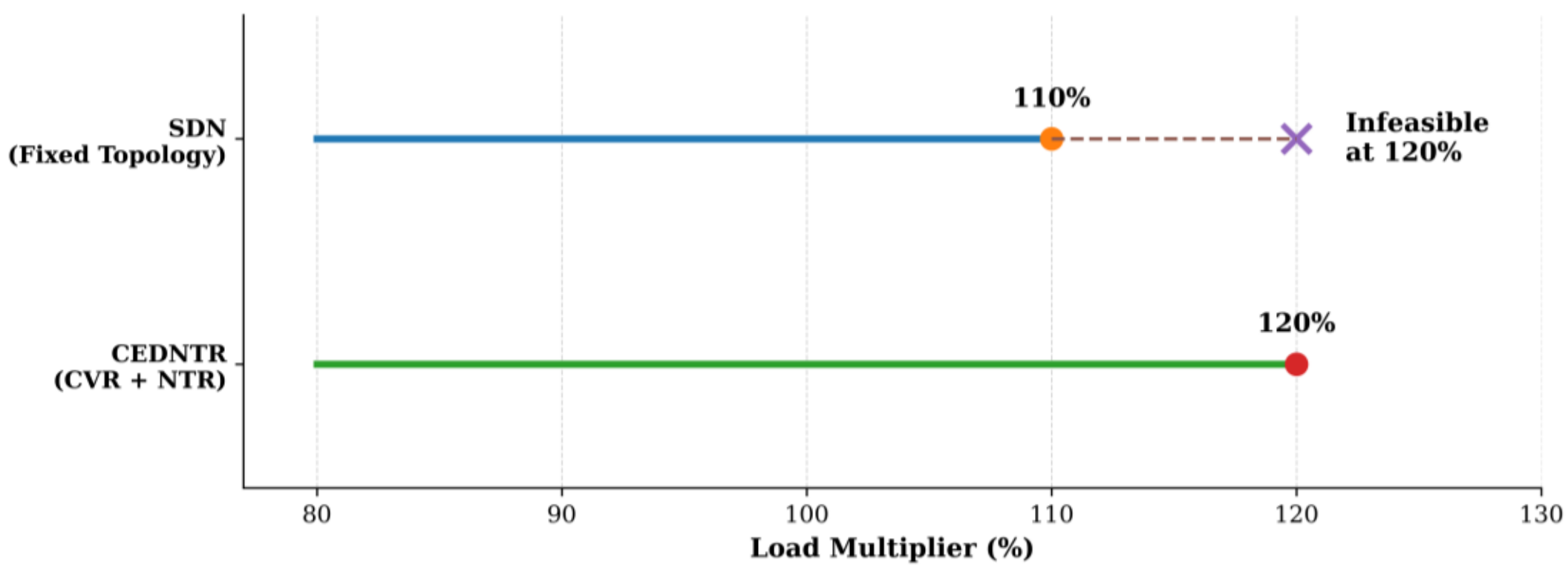

Fig. 13. Highest tested feasible loading level for SDN and CEDNTR.

Fig. 13 summarizes the demonstrated loading capability. Relative to the SDN baseline, CEDNTR demonstrates feasibility at the tested 120% loading level, compared with the highest tested feasible level of 110% for SDN.

## 5. Conclusion

This paper presented a unified day-ahead MISOCP framework for coordinating CVR and NTR in active distribution networks with ZIP loads, DERs, BESSs, and multiple substations. The results show that CVR and NTR provide distinct but complementary benefits that depend on load composition, DER portfolio scaling, feeder topology, voltage margins, and source capacity constraints. On the IEEE 33 bus system under representative mixed ZIP loading, NTR reduces active energy loss by approximately 2.50%, while coordinated CEDNTR achieves a 2.91% active energy loss reduction relative to SDN and enables a 2.46% CVR induced active load reduction relative to SDNTR. As load voltage sensitivity increases, CVR becomes substantially more effective, with CEDNTR achieving approximately 17.65% loss reduction relative to SDN and 14.61% CVR induced active load reduction relative to SDNTR under constant impedance loading. DER sensitivity further shows a nonmonotonic combined CEDNTR response, with the strongest load reduction occurring at the investigated 20% penetration level. On the IEEE 123 bus system, NTR reduces active energy losses by approximately 19.3%, while CVR primarily reduces voltage dependent demand by 2.3 to 2.4%. CEDNTR also remains feasible at 120% loading while SDN becomes infeasible, demonstrating improved loading robustness. Overall, coordinated CVR and NTR enable operating points that may not be captured when the two mechanisms are considered independently.